\documentclass[%
reprint,
superscriptaddress,
 amsmath,amssymb,
 aps,
 prapplied,
]{revtex4-1}

\usepackage{graphicx}
\usepackage{dcolumn}
\usepackage{bm}
\usepackage{hyperref}

\usepackage[caption=false]{subfig}
\usepackage{siunitx}
\usepackage{chemformula}
\usepackage{amsthm}
\usepackage{physics}
\usepackage{xcolor}

\begin{document}

\preprint{APS/123-QED}

\title{Current crowding in nanoscale superconductors within the Ginzburg-Landau model}


\author{Mattias J\"onsson}
\affiliation{%
 Department of Physics, KTH Royal Institute of Technology\\
 AlbaNova University Center, SE 106 91 Stockholm, Sweden
}
\author{Robert Vedin}
\affiliation{%
 Department of Physics, KTH Royal Institute of Technology\\
 AlbaNova University Center, SE 106 91 Stockholm, Sweden
}
\author{Samuel Gyger}
\affiliation{%
 Department of Applied Physics, KTH Royal Institute of Technology\\
 AlbaNova University Center, SE 106 91 Stockholm, Sweden
}
\author{James A. Sutton}
\affiliation{%
 Department of Applied Physics, KTH Royal Institute of Technology\\
 AlbaNova University Center, SE 106 91 Stockholm, Sweden
}
\author{Stephan Steinhauer}
\affiliation{%
 Department of Applied Physics, KTH Royal Institute of Technology\\
 AlbaNova University Center, SE 106 91 Stockholm, Sweden
}
\author{Val Zwiller}
\affiliation{%
 Department of Applied Physics, KTH Royal Institute of Technology\\
 AlbaNova University Center, SE 106 91 Stockholm, Sweden
}
\author{Mats Wallin}
\affiliation{%
 Department of Physics, KTH Royal Institute of Technology\\
 AlbaNova University Center, SE 106 91 Stockholm, Sweden
}
\author{Jack Lidmar}
\email{jlidmar@kth.se}
\affiliation{%
 Department of Physics, KTH Royal Institute of Technology\\
 AlbaNova University Center, SE 106 91 Stockholm, Sweden
}




\date{\today}

\begin{abstract}
The current density in a superconductor with turnarounds or constrictions is non-uniform due to a geometrical current crowding effect. This effect reduces the critical current in the superconducting structure compared to a straight segment and is of importance when designing superconducting devices. We investigate the current crowding effect in numerical simulations within the generalized time-dependent Ginzburg-Landau (\mbox{GTDGL}) model. The results are validated experimentally by measuring the magnetic field dependence of the critical current in superconducting nanowire structures, similar to those employed in single-photon detector devices. Comparing the results with London theory, we conclude that the reduction in critical current is significantly smaller in the \mbox{GTDGL} model. This difference is attributed to the current redistribution effect, which reduces the current density in weak points of the superconductor and counteracts the current crowding effect. We numerically investigate the effect of fill factor on the critical current in a meander and conclude that the reduction of critical current is low enough to justify fill factors higher than $\SI{33}{\percent}$ for applications where detection efficiency is critical. Finally, we propose a novel meander design which can combine high fill factor and low current crowding.
\end{abstract}

\maketitle


\section{\label{sec:introduction}INTRODUCTION}
Mesoscale and nanoscale superconductors play a central role in contemporary device configurations applied in various research fields. Inhomogeneous current density distributions, often termed current crowding, can occur at geometrical features such as corners, bends and constrictions, and severely impact the device properties of microwave resonators, filters and waveguides. Another prominent example is superconducting nanowire structures, which is the crucial building block for realizing devices capable of detecting light at the single-photon level \cite{Goltsman2001}. Due to their outstanding performance in terms of detection efficiency, time resolution and low intrinsic dark count rates, superconducting nanowire single-photon detectors (\mbox{SNSPDs}) have found wide-spread applications in, for instance, quantum optics, light detection and ranging (\mbox{LIDAR}), biological imaging and astronomy \cite{Hadfield2016,Holzman2019,Zadeh2021}. Furthermore, \mbox{SNSPDs} can be embedded in photonic integrated circuits \cite{Ferrari2018,Moody2021}, which allows for miniaturized implementations of complex quantum information processing architectures. In general, the detection mechanism of \mbox{SNSPDs} \cite{Engel2015} relies on operating the device relatively close to its critical current to enable the transduction of single photon absorption events into an electrical signal at high quantum efficiency. Hence, for further optimization of superconducting detectors and, more generally, of superconducting device properties, it is necessary to understand and mitigate the impact of current crowding, as the latter most commonly limits critical current and device functionality.

Current crowding effects in superconducting nanowire structures have been previously reported for the case of nanowire meanders with high fill factors \cite{Yang2009} and attributed to inhomogeneous current distributions at the bends within the London model \cite{Clem2011b}. Moreover, the importance of the geometry on the critical current \cite{Hortensius2012,Henrich2012} and on the observed dark counts during \mbox{SNSPD} operation \cite{Akhlaghi2012} has been highlighted. The application of magnetic fields has been proposed for probing and counteracting critical current reduction due to crowding at nanowire bends using time-dependent Ginzburg-Landau simulations \cite{Clem2012}. Several experiments have
demonstrated an asymmetric critical current with respect to a reversal of the direction of an applied magnetic field perpendicular to the plane of the nanowire device \cite{Adami2013,Ilin2014,Charaev2016}. To provide guidance for further device design optimization, it is essential to benchmark different theoretical models related to inhomogeneous supercurrent distributions, explore their limitations, and validate them with experimental data.

This paper focuses on the nanowire hairpin, the basic building block of superconducting meanders, made up from a single $\SI{180}{\degree}$ bend connected to two leads. The generalized time-dependent Ginzburg-Landau (\mbox{GTDGL}) model is used to investigate the effect of the inner bend geometry on the critical current. To verify the simulations, the effect of an applied magnetic field on the critical current of such superconducting nanowire structures is experimentally measured. Our results within the GTDGL model differ from those obtained with the London model, demonstrating that the latter significantly overestimates the impact of current crowding on the critical current in nanoscale superconductors. Hence, our results provide guidance for geometrical device design to achieve nanostructures with current carrying capabilities close to the material limits.

The paper is organized as follows. In Sec.~\ref{sec:current-crowding-in-the-ginzburg-landau-model} the current crowding effect in the Ginzburg-Landau model is studied for the simple case of an annulus disk geometry and compared with the London model. In Sec.~\ref{sec:theoretical-model} the theoretical model and the numerical methods used to simulate various hairpin geometries are described. In Sec.~\ref{sec:experimental-setup} the fabrication and experimental setup are presented. In Sec.~\ref{sec:critical-current-dependence-on-external-magnetic-field} simulations and experimental results for the critical current dependency on the applied magnetic field are presented. In Sec.~\ref{sec:current-crowding-in-gtdgl-vs-potential-flow} the current crowding effect in the \mbox{GTDGL} and London models are compared for various hairpin geometries. In Sec.~\ref{sec:critical-current-dependence-on-fill-factor} the critical current dependency on fill factor is studied, and in Sec.~\ref{sec:L-bend} we suggest a novel meander geometry to mitigate current crowding. Finally, Sec.~\ref{sec:conclusions} gives a summary and discussion of the results.

\section{\label{sec:current-crowding-in-the-ginzburg-landau-model}CURRENT CROWDING IN THE LONDON AND GINZBURG-LANDAU MODEL}

\begin{figure}[t]
    \centering
    \includegraphics[width=\linewidth]{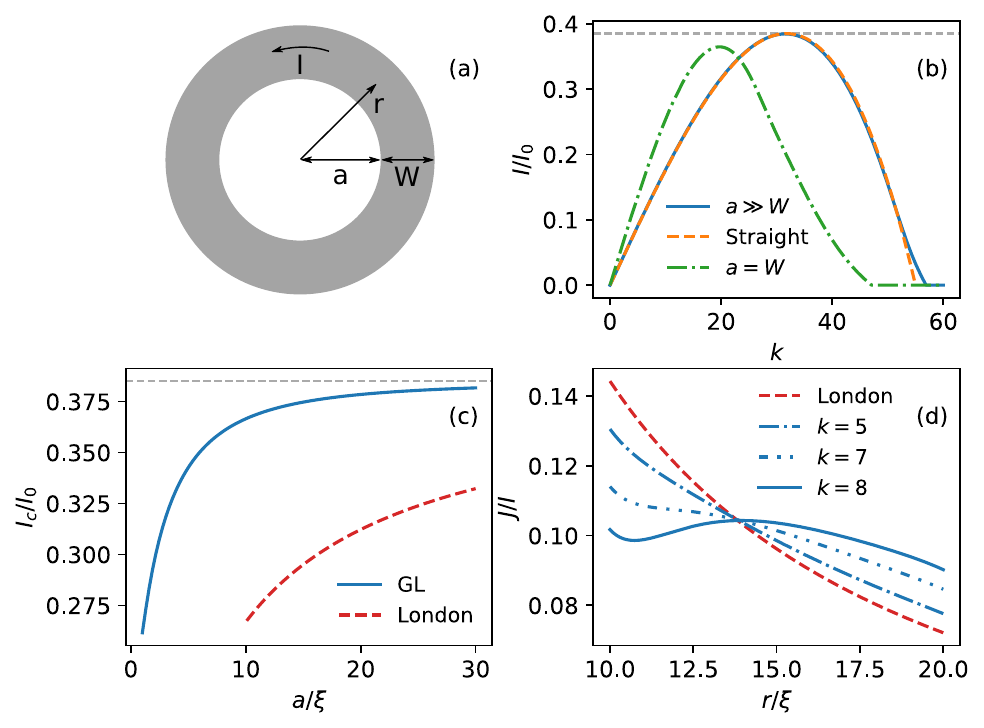}
    \caption{Current crowding effect in the Ginzburg-Landau model for an annulus area between concentric circles. (a) Geometry with inner radius $a$ and width $W$ used in the calculation. A current $I$ is induced into the superconductor (gray) by setting $k$ (see discussion below Eq. \eqref{eq:cylindrical-ginzburg-landau}). (b) Current as a function of applied flux (blue curve) for inner radius $a = 50 \xi$ and width $W = 10 \xi$. In the limit $a \gg W$ the geometry may be approximated with a straight superconductor and the analytical result from Eq.~\eqref{eq:j-vs-k-tinkham} provides excellent agreement (dashed orange curve). For $a = 25 \xi$ and $W = 25 \xi$ (dash-dotted green curve) the current crowding effect is non-negligible and the critical current is reduced below $I_{\text{dep}} = 0.385 I_0$. (c) Critical current as function of the inner radius. The Ginzburg-Landau model (blue curve) predicts significantly higher critical currents compared to the London model (dashed red curve). (d) Current density along a radial cross section for the London model (dashed red curve) and the Ginzburg-Landau model (blue curves: $k = 5$ dashed-dotted, $k = 7$ dashed-double-dotted, and $k = 8$ solid).}
    \label{fig:concentric-circles}
\end{figure}

The supercurrent density is given by $\mathbf{J} \propto \abs{\psi}^2 (\nabla \theta - \mathbf{A})$, where $\psi = \abs{\psi} e^{i\theta}$ is the complex order parameter. In the London limit the amplitude $\abs{\psi}$ is assumed constant, and in the absence of vortices the phase $\theta$ is single valued. If magnetic field fluctuations are neglected, formally by letting the London penetration depth $\lambda \to \infty$, the current pattern is described by an incompressible potential flow $\mathbf{J} \propto \nabla \theta$, with $\nabla \cdot \mathbf{J} \propto \nabla^2 \theta = 0$. 

The current crowding effect has been thoroughly studied in Ref.~\cite{Clem2011b} in the potential flow limit using the London model. By using conformal mapping techniques to compute the energy barrier for vortex crossings, it was concluded that a turnaround or a constriction results in a reduced critical current compared to a straight superconductor. Computing the vortex barrier to find the critical current for a general superconducting geometry is difficult, but it is possible to simplify the calculations when the smallest radius-of-curvature $R$ is much larger than the coherence length $\xi$, which is often the case in practical applications. In this case, the critical current is obtained when the local critical current density is  reached somewhere in the superconductor.  It is therefore sufficient to solve the Laplace equation to find the critical current.

In Ref.~\cite{Clem2011b} it is shown that most $\SI{180}{\degree}$ turnarounds result in a significant critical current reduction and in general sharper turnarounds tend to result in a larger reduction. It is also shown that there exists an optimal turnaround geometry that does not reduce the critical current compared to a straight superconductor, within London theory. The inner boundary of the optimal bend is for $x \leq \qty(2 W / \pi) \ln 2$ parameterized as $y_\pm(x) = \pm \qty(2 W / \pi)\cos^{-1}\qty[\exp{\pi x / 2W} / 2]$ (see Fig.~10 in Ref.~\cite{Clem2011b}), which takes the limiting value $y_\pm = \pm W$ far from the bend. This means that two meander lines must be separated by a distance of $2W$ in order to use this design. The optimal turnaround geometry is thus ideal in terms of current crowding. A drawback is that the fill factor, defined as the fraction of area covered by the superconducting film, is limited to $\SI{33}{\percent}$.

The validity of the London model is limited to small currents well below $I_c$. Currents close to the critical current are outside this regime and hence results from the London model do not necessarily apply. This motivates study of the current crowding effect in the Ginzburg-Landau model, which remains valid close to the critical current.

To clearly demonstrate the differences in the current crowding effect in the Ginzburg-Landau model compared to the London model we consider a superconducting disk geometry given by the annulus between two concentric circles. The width is $W$, thickness $d$, inner radius $a$, and coherence length $\xi$, as depicted in Fig.~\ref{fig:concentric-circles} (a). The cylindrical symmetry of this geometry may be used to simplify the time-independent Ginzburg-Landau model
\begin{equation}
    0 = \qty(\nabla - i \mathbf{A})^2 \psi + (1 - \abs{\psi}^2) \psi,
\end{equation}
into a non-linear ordinary differential equation by assuming that the superconducting order parameter $\psi = \abs{\psi}(r) e^{i \theta(\varphi)}$, where $r$ and $\varphi$ are cylindrical coordinates. Dimensionless units are used, as defined in Sec.~\ref{sec:theoretical-model}. The resulting ordinary differential equation is
\begin{equation}
    0 = r^2 \pdv[2]{\abs{\psi}}{r} + r \pdv{\abs{\psi}}{r} + (r^2 - k^2) \abs{\psi} - r^2 \abs{\psi}^3,
    \label{eq:cylindrical-ginzburg-landau}
\end{equation}
where $\theta(\varphi) = n \varphi$. A circulating current may be induced by threading a thin flux tube through the ring, or by setting a nonzero winding number $n$ of the phase, so that $k = n - \Phi/\Phi_0$, where $\Phi_0$ is the flux quantum. Isolating boundary conditions are used on the superconductor to vacuum interface, i.e. $\partial \abs{\psi} / \partial r = 0$. From the numerical solution of Eq.~\eqref{eq:cylindrical-ginzburg-landau} the supercurrent density is obtained as
$J(r) \propto \abs{\psi}^2 (\nabla \theta - \mathbf{A})= k\abs{\psi}^2/r$.

Figure~\ref{fig:concentric-circles} (b) shows the calculated current $I = d \int_a^{a+W} J(r) \mathrm{d}r$ in units of $I_0 = \hbar W d / 2 \mu_0 e \lambda^2 \xi$ as a function of $k$. As seen there is a maximal current that can be sustained by a cylindrically symmetric solution. Higher currents necessarily break the symmetry by the nucleation of vortices that start to cross the wire and lead to dissipation in the system. For a geometry with large inner radius compared to the width $a / W \gg 1$, the critical current equals the depairing current $I_{\text{dep}}  = 2 I_0 / \sqrt{27} = 0.385 I_0$ and the curve is well approximated by the formula for a straight superconductor~\cite{tinkam}
\begin{equation}
    I = I_0 \frac{k}{k_0} \qty(1 - \frac{k^2}{k_0^2}),
    \label{eq:j-vs-k-tinkham}
\end{equation}
where the superfluid velocity $v_s \propto k$ and
$k_0$ is a normalization factor that sets the maximal current at $k = k_0 / \sqrt{3}$. For a smaller ratio of $a / W$ the critical current is less than the deparing current $I_{\text{dep}}$, and thus demonstrates a non-negligible current crowding effect.

In Fig.~\ref{fig:concentric-circles} (c) results for the critical current (normalized to $I_0$) as a function of the inner radius are compared for the Ginzburg-Landau and London models. In general the critical current is significantly higher in the Ginzburg-Landau model compared to the London model, which shows that the current crowding effect is less pronounced than previously reported in the literature. This qualitative difference between the models is caused by the current redistribution effect, which is shown in Fig.~\ref{fig:concentric-circles} (d). For sufficiently large $k$ the current density near the inner boundary is large enough to cause a significant suppression of the superconducting order parameter $\abs{\psi}$. This suppression in turn limits the current density locally, which forces the current to redistribute. This results in a flatter current profile with smaller peak values, which increases the critical current and therefore weakens the current crowding effect compared to the London limit.

\section{\label{sec:theoretical-model}THEORETICAL MODEL}

\begin{figure}[t]
    \centering
    \includegraphics[width=\linewidth]{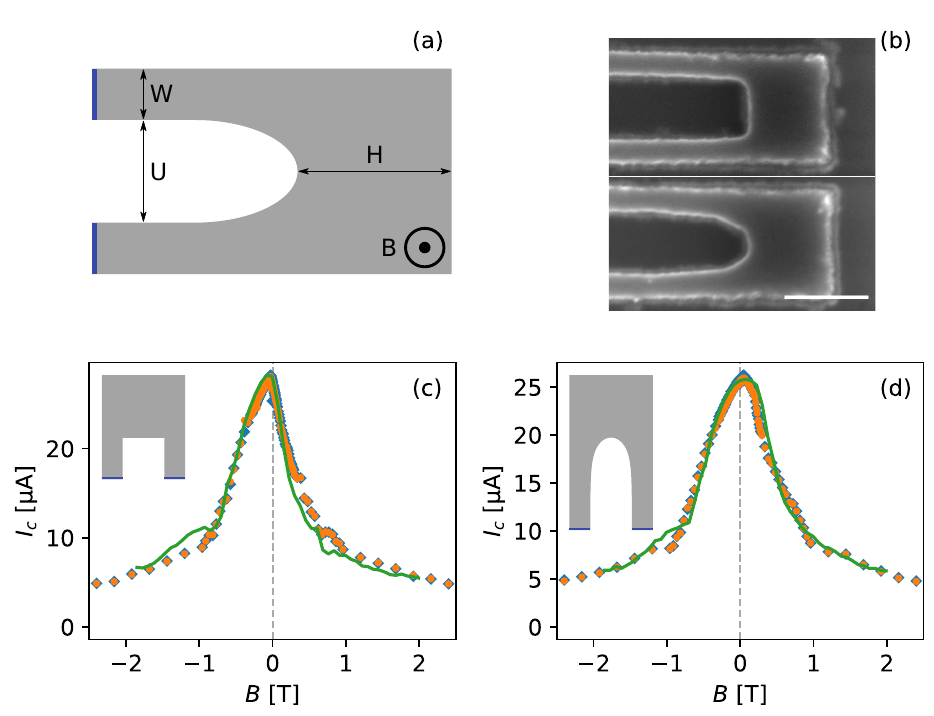}
    \caption{(a) A typical turnaround used in the simulations. A uniformly distributed current density $J$ is injected through the lower metal contact (blue) and the current flows through the superconductor (gray) to the upper metal contact. The leg width is $W = 14 \xi$, the head width $H = 3W$, and $U$ varies based on the fill factor. (b) SEM image of a hairpin device with square and optimal turnaround. The scalebar is \SI{200}{\nano\meter}. (c -- d) Experimentally measured critical current as a function of magnetic field (blue diamond) fitted with simulation data (green curve) for a $\SI{33}{\percent}$ fill factor square (c) and optimal (d) turnaround (shown in the insets). Experimental data for reversed current direction and reversed magnetic field (orange circle) overlaps the positive current data (blue diamond).}
    \label{fig:experimental-fit}
\end{figure}

To study the current crowding effect in a general geometry we consider a superconductor with thickness $d$ much smaller than the London penetration depth $\lambda$, i.e. $d \ll \lambda$, and width $W$ much smaller than the Pearl length $\Lambda = 2 \lambda^2 / d$, i.e. $W \ll \Lambda$. In this regime the magnetic field from the current passing through the superconductor has little influence on the magnetic field dynamics \cite{Clem2011b} and it is justified to make the approximation that the magnetic field $B$ is constant, normal to the superconducting surface and equal to the applied magnetic field.

To simulate the superconductor we employ the \mbox{GTDGL} model \cite{Kramer1978TheoryFilaments, Watts-Tobin1981NonequilibriumTc, Kopnin}
\begin{equation}
\begin{split}
    \frac{u}{\sqrt{1 + \gamma^2 \abs{\psi}^2}} \qty(\pdv{t} + i \mu + \frac{\gamma^2}{2} \pdv{\abs{\psi}^2}{t}) \psi =\\
    \qty( 1 - \abs{\psi}^2 ) \psi + \qty(\nabla - i \mathbf{A})^2 \psi,
    \label{eq:gtdgl}
\end{split}
\end{equation}
\begin{equation}
    \nabla^2 \mu = \nabla \cdot \Im{\psi^* (\nabla - i \mathbf{A}) \psi},
    \label{eq:gtdgl-current-conservation}
\end{equation}
\begin{equation}
    \mathbf{A} = \frac{-By}{2} \mathbf{\hat{x}} + \frac{Bx}{2} \mathbf{\hat{y}},
    \label{eq:gtdgl-vector-potential}
\end{equation}
where $\psi$ is the superconducting order parameter, $\mathbf{A}$ is the vector potential, and $\mu$ is the electric potential. We make a conventional choice for the material parameter $u = 5.79$ \cite{Kramer1978TheoryFilaments} and assume $\gamma = 10$. The dimensionless units are chosen such that distances are measured in the coherence length $\xi$, time is measured in $\tau = \mu_0 \sigma \lambda^2$ where $\sigma$ is the conductivity, voltages are measured in $v_0 = \hbar / 2 e \mu_0 \sigma \lambda^2$, currents are measured in $J_0 = \hbar / 2 \mu_0 e \lambda^2 \xi$, and magnetic fields are measured in $B_{c2} = \hbar / 2 e \xi^2$. Equation~\eqref{eq:gtdgl-vector-potential} is a result of the approximation that the magnetic field is constant and equal to the applied magnetic field $B$.

We simulate the \mbox{GTDGL} model for geometries such as the turnaround presented in Fig.~\ref{fig:experimental-fit} (a). The superconductor is coupled to two metal terminals with opposite normal current densities $J_{\text{ext}}$, that control the current flowing through the superconductor. This is modelled by applying the boundary conditions $\psi = 0$ and $\mathbf{\hat{n}} \cdot \nabla \mu = J_{\text{ext}}$ on the superconductor to metal interfaces. The remaining boundaries are superconducting to vacuum interfaces and the boundary conditions are given by $\mathbf{\hat{n}} \cdot (\nabla - i \mathbf{A}) \psi = 0$ and $\mathbf{\hat{n}} \cdot \nabla \mu = 0$ to ensure no supercurrent or normal current flows across the boundary.

We solve the \mbox{GTDGL} equations \eqref{eq:gtdgl} and \eqref{eq:gtdgl-current-conservation} using a finite volume method on an unstructured Delaunay triangulation \cite{Mathematics2021Analysis1049-10, Du2004NumericalSphere}, where the edge lengths in the triangles are at most one half coherence length. Link variables are used for the magnetic vector potential $\mathbf{A}$ by introducing $U = \exp{-i \int \mathbf{A} \cdot \dd{\mathbf{l}}}$ and rewriting the covariant derivative as
\begin{equation}
    \qty(\nabla - i \mathbf{A})\psi = U^*\nabla \qty(U \psi).
\end{equation}
This formulation of the covariant derivative preserves the gauge invariance when discretized on a lattice \cite{GROPP1996254}. The electrical scalar potential $\mu$ is treated similarly for the covariant time derivative. The discretized form of Eq.~\eqref{eq:gtdgl} is solved using a semi-implicit Euler method and Eq.~\eqref{eq:gtdgl-current-conservation} is solved with a sparse LU factorization.

\section{\label{sec:experimental-setup}EXPERIMENTAL SETUP}
Device fabrication was performed on silicon substrates covered with thermal \ch{SiO2}, relying on a room-temperature reactive sputtering process of \ch{NbTiN} with stoichiometry optimized for single-photon detectors \cite{Zichi2019}. The \ch{NbTiN} thin films (thickness $\SI{9}{\nano\meter}$) were patterned into nanowire structures via electron beam lithography and reactive ion etching. The \mbox{SNSPDs} consist of hairpins with a width of $W=\SI{70}{\nano\meter}$ and a fill factor of \SI{33}{\percent} ($U=\SI{140}{\nano\meter}$). The leads are \SI{50}{\micro\meter} long and connected to a lumped-element inductor to avoid latching. The detectors were mounted inside a closed-cycle dipstick system (Attodry 2100) with \ch{He} exchange gas at temperatures below $\SI{2}{\kelvin}$ inside a uni-axial field created by a superconducting magnet and connected to coaxial cables leading to room temperature. The \mbox{SNSPD} were operated using commercial control electronics (Single Quantum Atlas) containing a bias-tee and a two-stage room temperature amplifier. The measurement sweeps were controlled using Python and critical currents were extracted by fitting a step-function using lmfit \cite{10.5281/zenodo.11813}.
The measurement configuration is schematically depicted in Fig.~\ref{fig:experimental-fit} (a), whereas representative scanning electron micrographs of the investigated devices are presented in Fig.~\ref{fig:experimental-fit} (b). This device design is immediately relevant for waveguide-integrated superconducting single-photon detectors employed in nanophotonic circuits \cite{Gyger2021}.

\section{\label{sec:results}RESULTS}

\subsection{\label{sec:critical-current-dependence-on-external-magnetic-field}CRITICAL CURRENT DEPENDENCY ON APPLIED MAGNETIC FIELD}

\begin{figure}[t]
    \centering
    \includegraphics[width=\linewidth]{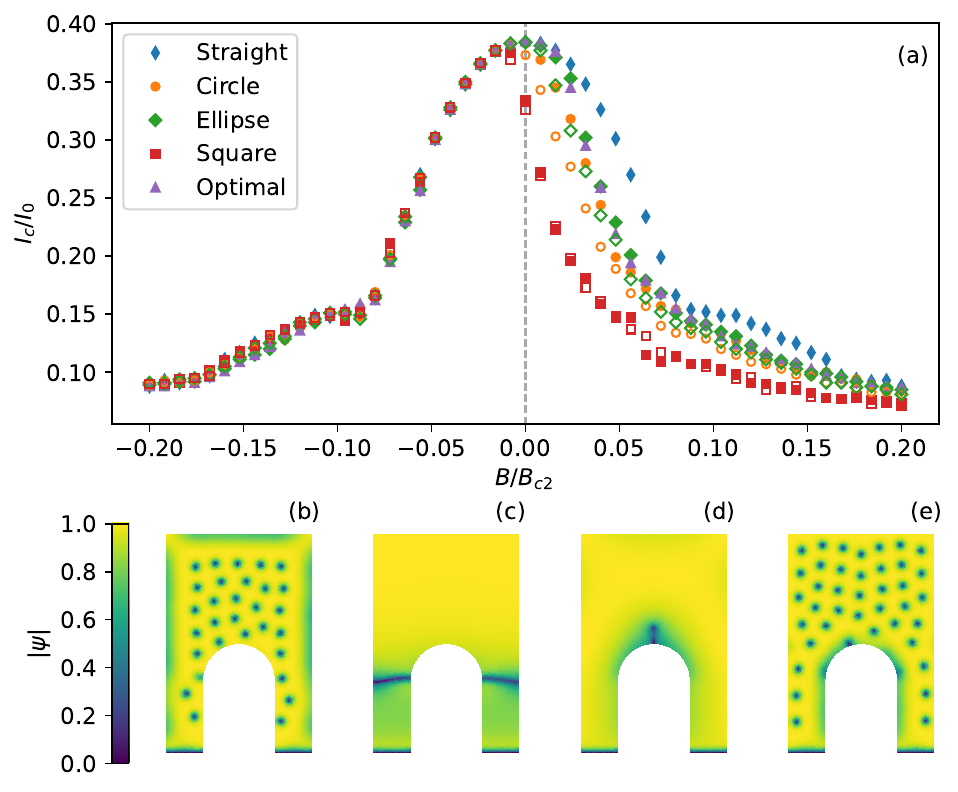}
    \caption{Simulation results from the \mbox{GTDGL} model. (a) The simulated critical current as a function of the applied magnetic field for geometries with $\SI{33}{\percent}$ fill factor (filled markers) and $\SI{50}{\percent}$ fill factor (hollow markers). A straight superconducting strip (blue diamonds) is added as a current crowding free reference. (b -- e) Superconducting order parameter showing the breakdown mechanics slightly above $I_c$. (b) $B / B_{c2} = -0.1$. Vortices nucleate on the outer boundary and move across the nanowire. (c) $B / B_{c2} = 0$. Vortices nucleate at the start of the inner bend and create a phase-slip line. (d) $B / B_{c2} = 0.025$. Vortices nucleate at the tip of the turnaround. (e) $B / B_{c2} = 0.1$. Vortices nucleate on the inner bend and leave the superconductor on the outer boundary.}
    \label{fig:ic-vs-b-simulation-results}
\end{figure}

An applied perpendicular magnetic field will, depending on direction, either increase or decrease the amount of current crowding at the inner bend of a turn. This will then change the energy barrier for vortex crossings and lead to a lower or higher critical current of the structure, and will show up as an asymmetry in $I_c(B)$ about zero field.
For practical applications with both left and right bends, a homogeneous magnetic field could reduce the current crowding only at one type of bend.
Still, from an experimental point, this is a useful diagnostic of current crowding in hairpin geometries as studied here, since it relies on comparing the critical current in the same device at different fields, rather than different devices with varying geometries, which are susceptible to sample-to-sample variations.

The critical current dependence on the applied magnetic field is investigated in both experiment and simulation by extracting critical currents from IV curves for different constant applied fields. The results from experimental devices with fitted simulation data for the square turnaround and the optimal turnaround are presented in Fig.~\ref{fig:experimental-fit} (c) and (d), respectively. To fit the simulation data dimensions were reintroduced to the current by multiplying with $I_0 = W d J_0$ and to the magnetic field by multiplying with $B_{c2}$. We also added an empirical constant offset $B_{\text{off}}$ to the simulated magnetic field. The best fit corresponds to $B_{c2} = \SI{10}{\tesla}$, $B_{\text{off}} = \SI{70}{\milli\tesla}$, $I_0 = \SI{75}{\micro\ampere}$ for the device with square turnaround, and $I_0 = \SI{67}{\micro\ampere}$ for the optimal turnaround. These values correspond to the estimated parameters $\xi \approx \SI{5.7}{\nano\meter}$ and $\lambda \approx \SI{600}{\nano\meter}$ for $T \approx \SI{2}{\kelvin}$. Extrapolation to $T = 0$ using BCS theory gives $\xi(0) \approx \SI{5.1}{\nano\meter}$ and $\lambda(0) \approx \SI{550}{\nano\meter}$ \cite{tinkam}. These are consistent with previously reported values for \ch{NbTiN} films~\cite{Sidorova2021, Thoen2017SuperconductingWafer}. The assumed magnetic offset $B_{\text{off}}$ moves the simulation data peak curve slightly closer to zero magnetic field and is needed for a good fit. The origin of the offset is not known. However, it is unlikely that it is due to the presence of a constant magnetic field background as shown by the overlapping data points in Fig.~\ref{fig:experimental-fit} (c -- d). In the literature a stronger peak asymmetry has been reported \cite{Adami2013, Ilin2014, Charaev2016}, which motivates further study.

The simulation results in Fig.~\ref{fig:experimental-fit} (c -- d) are in good agreement with the experimental data, showing a larger asymmetry for the geometry with sharp corners. In a region around $B = \SI{-1}{\tesla}$ the simulations display a bump which is not present in the experimental results. The bump seen in the simulation data has previously been reported in Refs. \cite{Clem2012, Vodolazov2013Vortex-inducedFilms, Ilin2014} and is attributed to formation of a vortex lattice. 
The precise shape and location of bumps in this regime is likely to depend on the detailed geometry of the device.
One possible explanation for the missing bumps in the experimental data shown in Fig.~\ref{fig:experimental-fit} (c -- d) is that vortices move more freely for non-zero temperatures. Another difference between experiment and simulation is present for the square turnaround when a positive magnetic field is applied, i.e. the simulation predicts a smaller critical current compared to experiment. This difference is likely explained by the $\SI{90}{\degree}$ corners being rounded in the experimental device, as seen in the upper SEM image in Fig.~\ref{fig:experimental-fit} (b).

Figure~\ref{fig:ic-vs-b-simulation-results} (a) shows simulated critical current vs magnetic field for different geometries: straight, circle, ellipse with aspect ratio $1:2$, square, and optimal according to London theory. As expected from the symmetry of the straight superconducting strip, its critical current is symmetric around $B = 0$. For the remaining geometries the critical current is asymmetric around $B = 0$. As discussed this asymmetry is a clear sign of current crowding being influenced by the screening currents induced by the applied magnetic field.

For sufficiently strong negative magnetic fields, the critical current of the turnaround geometries coincide with the straight geometry, which is explained by the weakest point being located on the legs. This is shown in Fig.~\ref{fig:ic-vs-b-simulation-results} (b), where it is seen that the first non-stationary vortices nucleate on the outer boundary of the legs and flow inward towards the inner boundary. For zero magnetic field, all smooth geometries at $\SI{33}{\percent}$ fill factor have negligible reduction in critical current, which is consistent with Fig.~\ref{fig:ic-vs-b-simulation-results} (c) where the weakest point is in the start of the inner bend. For small positive magnetic fields around $B / B_{c2} = 0.05$, the critical current is reduced compared to the straight geometry, which shows that the magnetic field aggravates the current crowding effect. This is seen in Fig.~\ref{fig:ic-vs-b-simulation-results} (d), where the weakest point has moved to the tip of the turnaround as a result of current crowding. For sufficiently strong positive magnetic field, the critical current for the turnarounds approach the value for the straight geometry. In Fig.~\ref{fig:ic-vs-b-simulation-results} (e) non-stationary vortices nucleate everywhere on the inner boundary, which suggests that the inner bend is less limiting for the critical current.

\subsection{\label{sec:current-crowding-in-gtdgl-vs-potential-flow}CURRENT CROWDING IN GTDGL VS LONDON}

In Sec.~\ref{sec:current-crowding-in-the-ginzburg-landau-model} we showed that the current crowding effect in the Ginzburg-Landau model is qualitatively different compared to the London model when considering a system of concentric circles. The difference between the models is that Ginzburg-Landau takes the suppression of the superconducting order parameter into account, which results in a redistribution of the current density. The current redistribution effect depends on the turnaround geometry and we therefore investigate different geometries.

In \mbox{GTDGL} model, we define the critical current as the current at which vortices start to flow across the wire, since the vortex motion will generate a voltage. In the London model, we define the critical current at which the vortex barrier for crossings is zero. For a bend with radius of curvature $R$ much larger than $\xi$, this will coincide with the current when the critical current density is locally reached somewhere \cite{Clem2011b}. For simplicity the comparisons is restricted to geometries such that $R \ge 10 \xi$ by scaling up the size if necessary \footnote{For the square geometries we use a rescaling factor such that the spacing between the legs is much larger than the coherence length and use Eq.~(100) in Ref.~\cite{Clem2011b} to compute the reduction factors.}.

\begin{figure}[t]
    \centering
    \includegraphics[width=\linewidth]{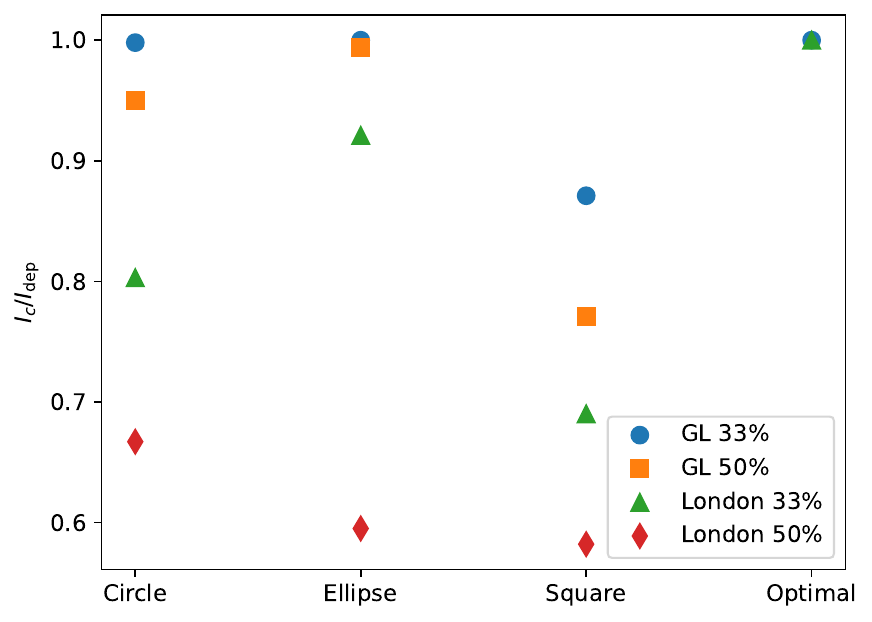}
    \caption{Reduction of the critical current compared to a straight superconductor. The reduction predicted by the \mbox{GTDGL} model with $\SI{33}{\percent}$ fill factor (blue circle) and $\SI{50}{\percent}$ fill factor (orange square) are compared with the London model for $\SI{33}{\percent}$ fill factor (green triangle) and $\SI{50}{\percent}$ fill factor (red diamond). All geometries have been rescaled such that the smallest radius of curvature $R \ge 10 \xi$, which allows for direct comparison between the \mbox{GTDGL} and London models without any extra compensation factors due to small geometries. For all geometries, except for the optimal, the London model predicts a significantly larger reduction of the critical current than the \mbox{GTDGL} model.}
    \label{fig:gtdgl-vs-potential-flow}
\end{figure}

In Fig.~\ref{fig:gtdgl-vs-potential-flow} the reduction of critical current is compared between the \mbox{GTDGL} model and the London model for the different geometries. All geometries, except the optimal geometry, display significantly smaller reduction in the \mbox{GTDGL} model compared to the London model. This shows that the London model overestimates the impact of current crowding and turnaround geometry. From the \mbox{GTDGL} simulations we may therefore conclude that for sufficiently wide heads (large $H$ in Fig.~\ref{fig:experimental-fit} (a)) the exact shape of a rounded inner bend does not significantly affect the critical current for fill factors up to $\SI{50}{\percent}$. Defects in the manufacturing process may be a larger contribution to the reduction than the geometrical form of the inner bend.

\subsection{\label{sec:critical-current-dependence-on-fill-factor}CRITICAL CURRENT DEPENDENCE ON FILL FACTOR}

\begin{figure}
    \centering
    \includegraphics[width=\linewidth]{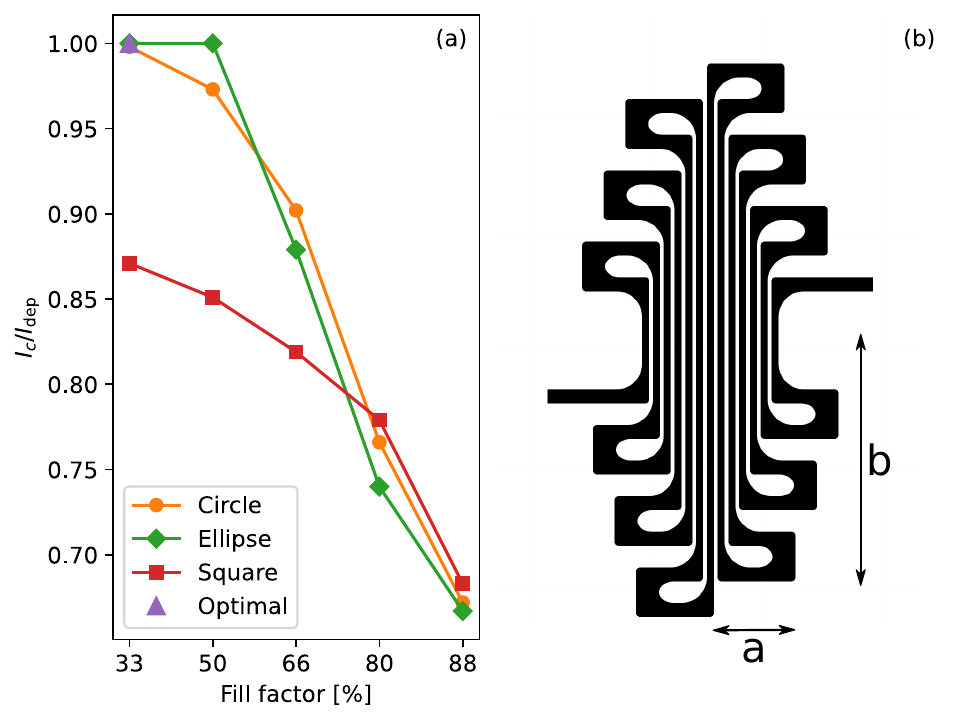}
    \caption{(a) Reduction of the critical current compared to a straight superconductor as a function of the fill factor. For $\SI{33}{\percent}$ fill factor the reduction of the critical current is negligible for all smooth turnarounds. The critical current is reduced as the fill factor increases and for $\SI{88}{\percent}$ fill factor all geometries converge to the same critical current. (b) An example of a meander with L-bends used to minimize the current crowding effect, while simultaneously allowing an arbitrary fill factor. The use of L-bends results in a changed aspect ratio with $a \neq b$.}
    \label{fig:ic-vs-ff}
\end{figure}

Figure~\ref{fig:ic-vs-ff} (a) shows the critical current reduction as a function of the fill factor for the \mbox{GTDGL} simulations. The fill factor is tuned for each geometry by changing the spacing between the legs $U$ while keeping $W = 14 \xi$ fixed (see Fig.~\ref{fig:experimental-fit} (a)) and rescaling the inner bend to leave the geometry unchanged. The critical current decreases as the fill factor increases, which is expected since the current crowding effect becomes stronger for sharper turnarounds. The numerical results for the critical current of the rounded geometries display a similar reduction as the experimental results in Ref.~\cite{Yang2009}. The latter were based on devices with square inner bends that tend to be rounded due to fabrication limitations which makes them more comparable to the rounded simulation geometries. Furthermore, the reduction in critical current is less than what is expected from London theory which has been seen experimentally in Refs. \cite{Korneeva2021InfluenceMeanders, Meng:20}

The shape of the inner bend plays a role in how quickly the critical current drops as the fill factor is increased. As a result, the best choice of inner bend may change as the fill factor increases. In the limiting case approaching $\SI{100}{\percent}$ fill factor it is clear that all radii-of-curvature are zero, all geometries have the same shape and the critical currents are equal.
Figure \ref{fig:ic-vs-ff} (a) can be used to determine a tradeoff between the optical design for high detection efficiency \cite{Anant2008} and the suppression of the critical current by varying the fill factor.

\subsection{\label{sec:L-bend}MEANDER WITH ARBITRARY FILL FACTOR AND LOW CURRENT CROWDING}

As shown in Sec.~\ref{sec:critical-current-dependence-on-fill-factor}, the \mbox{GTDGL} simulations indicate that the critical current of the structure is not as sensitive to the geometry as previously reported. However, for high fill factors the current crowding effect still leads to a considerable reduction of the critical current. In the case of a standard meander composed of straight lines connected with $\SI{180}{\degree}$ turnarounds, there is a tradeoff between having high critical current and a high fill factor. One possible solution is to increase the thickness of the superconducting film at the bends in order to locally reduce current density~\cite{Baghdadi2021EnhancingThickness}.

We propose a new meander design that eliminates the tradeoff between critical current and fill factor while avoiding the need for variable film thickness. The concept uses L-bends, which are turnarounds consisting of an extra $\SI{90}{\degree}$ turn before the $\SI{180}{\degree}$ turnaround, as depicted in Fig.~\ref{fig:ic-vs-ff} (b). This design allows the fill factor near the turnaround to be smaller than the fill factor of the light-detecting meander lines, which makes it possible to use an arbitrary fill factor while allowing the turnaround to be wider and have low fill factor. The downside of this design is that the meander will be elongated due to the extra space required for the L-bends and in general the ratio $b / a > 1$.

The ratio $b / a$ may for a large number of lines be approximated as a ratio between the pitch in the L-bend $P_B$ and the pitch in the meander lines $P_M$. Alternatively it can also be approximated using the fill factor of the L-bend $\mathrm{ff}_B$, the meander $\mathrm{ff}_M$ and the ratio between the width of the superconducting wire in the bend and in the meander line $k = W_B / W_M$. Thus
\begin{equation}
    \frac{b}{a} = \frac{P_B}{P_M} = k \frac{\mathrm{ff}_M}{\mathrm{ff}_B}.
\end{equation}
In order to estimate the elongation of a L-bend meander, we consider the limiting case where the meander lines have width $W_M = W$ and $\SI{100}{\percent}$ fill factor, while the L-bends have width $W_B = 2 W$ to reduce dark counts and allow for a $\SI{33}{\percent}$ fill factor. These dimensions gives $b = 6 a$ and the meander area would increase by about a factor of 3 to cover the same circular area as a traditional meander.

During the manuscript review process, the authors became aware of the successful experimental implementation of a similar design in high-fill-factor superconducting microwire single-photon detectors, resulting in high detection efficiency combined with low polarization sensitivity \cite{Reddy2022BroadbandDetectors}.

\section{\label{sec:conclusions}SUMMARY}

We used the \mbox{GTDGL} model to explore the current crowding effect in superconducting nanostructures.
Theoretical predictions are tested experimentally by comparing how the critical current varies with the applied magnetic field for two distinct nanowire turnaround geometries. In comparison to the previously used London model, we show that current crowding is substantially less pronounced in the \mbox{GTDGL} model. This finding is explained by the suppression of the superconducting order parameter near bends, which causes current redistribution close to sharp turns. Within London theory it is possible to design optimal turnarounds without current crowding, but this restricts the fill factor to $\SI{33}{\percent}$. Using the more accurate description provided by the \mbox{GTDGL} model it is possible to achieve less than one percent reduction for $\SI{50}{\percent}$ fill factor. We propose that this marked difference in predicted critical currents considerably loosens the design restrictions for a large range of superconducting nanodevices and circuits. Hence, our results pave the way towards devices with improved performance and current carrying capabilities approaching the material limits.

\section*{\label{sec:acknowledgments}Acknowledgments}

The authors acknowledge funding from the Knut and Alice Wallenberg Foundation through the grant Quantum Sensors.
S.G. acknowledges funding from the Swedish Research Council under Grant Agreement No. 2016-06122 (Optical Quantum Sensing).
S.S. and V.Z. acknowledge support from European Union's Horizon 2020 research and innovation program (FastGhost Project, Surquid Project, and aCryComm Project).
This research was conducted using the resources of High Performance Computing Center North (HPC2N),
provided by the Swedish National Infrastructure for Computing (SNIC), partially funded by the Swedish Research Council through grant agreement no. 2018-05973.


%

\end{document}